\documentclass[aps,twocolumn,groupedaddress]{revtex4}
\usepackage{graphicx}

\begin{document}
\bibliographystyle{apsrev}


\title{Specular reflection of matter waves from a rough mirror}



\author{V. Savalli, D. Stevens, J. Est\`eve, P.D. Featonby, 
V. Josse, N. Westbrook, C.I. Westbrook, A. Aspect}
\affiliation{Laboratoire Charles Fabry de l'Institut d'Optique, UMR 
8501 du CNRS,  
91403 Orsay CEDEX, France}


\date{Current version \today}

\begin{abstract}
We have made a high resolution study of the specularity of the atomic 
reflection 
from an 
evanescent wave mirror using velocity
selective Raman transitions. 
We have observed a double structure
in the velocity distribution after reflection: a peak consistent
with specular reflection and a diffuse reflection
pedestal, whose contribution decreases rapidly with increasing detuning.
The diffuse reflection is due to two distinct effects: spontaneous 
emission in the evanescent wave and a roughness in the evanescent
wave potential whose amplitude is 
smaller than the de Broglie wavelength of the 
reflected atoms.
    
\end{abstract}


\maketitle

%

Atomic mirrors are key components in the growing field of atom
optics, and have been intensively studied by several groups in recent
years~\cite{saba:99a,cognet:99b,lau99a,arnold:02a,landragin:96a,bongs:99,bloch:01,
kadio:01,voigt:00a}. 
For interferometric and lithographic applications, it is
particularly important to ensure that the reflection at the mirror is
specular, since diffuse scattering amounts to a loss of spatial
coherence and consequently to a reduction of fringe visibility or
focusing sharpness. Thus much interest has been devoted to measuring
and improving the roughness of atomic mirrors, using static or time
dependent magnetic fields~\cite{saba:99a,cognet:99b,lau99a,arnold:02a},
evanescent waves~\cite{landragin:96a} or other
techniques~\cite{bloch:01}.

Most of the above mentioned experiments consisted in 
preparing an initial
narrow velocity distribution, transverse to
the direction of incidence, and measuring the broadening of this
distribution due to the reflection. 
The effect of
the mirror was characterized by a single quantity, the rms increase
in the width of the velocity distribution attributable to the mirror.
This increase was sometimes also given in terms of an effective rms
angular deviation from a perfectly flat surface. 
In those experiments however, the resolution only allowed
a measurement of the broadening of the atomic velocity
distribution due the mirror.
In this paper we present an experiment using velocity
selective stimulated Raman transitions to prepare a very narrow
initial velocity distribution~\cite{Kasevich:91a}, 
and to measure the distribution after reflection. 
For the first time
we are able to resolve a final velocity distribution which is not
merely broadened, but qualitatively modified upon reflection.

Much in analogy with the reflection of light from an optical mirror,
we observe a double structure, with a narrow peak and a broad
pedestal. The narrow peak corresponds to specularly reflected atoms.
The broad pedestal corresponds to diffuse atomic reflection
which has two origins: spontaneous emission and mirror roughness.
If we assume that the roughness of the mirror can be viewed 
as a random process with variance $\sigma^{2}$
and a correlation length much
shorter than the mirror itself,
one readily finds ~\cite{Garcia:93a,Henkel:97b} 
that the fraction $S$ of specularly reflected atoms 
is given by $\exp -w$, reminiscent of the Debye-Waller factor, 
with $w=16\pi^{2} \sigma^{2}/\lambda_{\mathrm{dB}}^{2}$, where 
$\lambda_{\mathrm{dB}}$ is the de Broglie wavelength 
of the incident matter wave (8 nm in our experiment).
The presence of a significant specular peak implies 
an effective
\footnote{Explicit calculations relating the effective mirror roughness to
actual substrate roughness are given in \cite{Garcia:93a,Henkel:97b}} 
mirror roughness
smaller than $\lambda_{\mathrm{dB}}$.

We have
studied the ratio of these two components as a function of various
parameters and shown in particular that the diffuse component rapidly
decreases when the atomic evanescent wave detuning increases. These
observations also allow us to examine different possible mechanisms
for the diffuse reflection involving either 
mirror roughness or spontaneous emission.

In our experiment we use an evanescent wave mirror, identical to
the one described in Ref.~\cite{Cognet:98a}. 
We use a superpolished prism of TaFD30 glass \footnote{purchased from 
General
Optics, Moorpark California} (refractive index $n_{1}=1.869$). 
The rms surface roughness given by
the manufacturer is $0.07$~nm.
The input
and output faces for the laser beam which creates the evanescent
wave are coated with a broadband AR reflection coating. 
A Ti:S
laser of wavelength $\lambda_{L} = 2 \pi / k_{L} = 780$~nm generates
the evanescent wave with an incident angle $\theta_1 =
53^{\circ}$. The evanescent electric field thus has a decay
constant of $\kappa = k_L \sqrt{n_1^2\sin^2\theta_1-1} = 1.11 k_{L}$, 
and a
propagation vector of magnitude $k_{x} = k_{L} n_{1} \sin
\theta_{1} = 1.49 k_{L}$.
We have defined the $x$-axis to be along the
evanescent wave propagation direction. The Ti:S laser is TM (p)
polarized and its waist is about $0.9$~mm along the $x$ and $z$-axes
at the surface of the prism.

Our $^{85}$Rb magneto-optical trap (MOT) is similar to that described in
Ref.~\cite{Cognet:98a}. Every $1.5$~s, we collect about
$10^8$ atoms in the trap. By turning off the repumping
beam just before (after) the trapping beams, we prepare the atoms
in the $| 5S_{1/2},F= 2\rangle$ ($| 5S_{1/2},F= 3\rangle$) level.
The MOT is situated $20$~mm above the prism.

\begin{figure}
\includegraphics{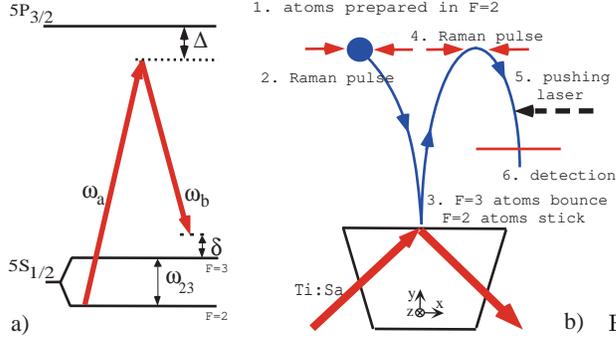}
\caption{ a) Level diagram for velocity selective stimulated Raman 
transitions. 
The frequencies $\omega_{a}$ and $\omega_{b}$ of the two
counterpropagating laser beams are separated by the 
$^{85}\mathrm{Rb}$ hyperfine 
frequency $\omega_{23}$ plus the Raman detuning $\delta$. 
b): Sequence used in our experiment to select and
analyze a narrow velocity class. 
The selection Raman pulse is applied as the atoms begin to fall 
toward 
the mirror 
and the analysis pulse comes at the top of their trajectory after the 
bounce. 
The Raman beams propagate at $43^\circ$ to the $x$-axis in the 
$xz$-plane.}
\label{fig.disporaman}
\end{figure}

Velocity selective Raman transitions between $F=2$ and $F=3$
hyperfine levels are induced by a pair of counter-propagating
laser beams detuned from the one photon atomic resonance by about
1 GHz ($\Delta$ in Fig.~\ref{fig.disporaman})~\cite{Kasevich:91a}.
Because of the Doppler effect, the resonance condition for a two
photon Raman transition depends on the atomic velocity and is
given by $\delta = 2 k(v+v_{\rm{rec}})$, where
$\delta=\omega_a-\omega_b-\omega_{23}$ is the detuning of the
Raman beams from the hyperfine transition, $v=\mathbf{v.k_a} /
k_a$ is the projection of the atomic velocity on the Raman beam
direction, and $v_{\rm{rec}}$ is the one photon recoil velocity of
Rb, 6 mm/s. By varying the detuning $\delta$, we can choose the
center of the velocity class that experiences a transition.

The two Raman beams are orthogonally linearly polarized and drive
the magnetic field independent Raman transition between
$|5S_{1/2}F=2,m_F=0 \rangle$ denoted $|2 \rangle$ and $
|5S_{1/2}F=3,m_F=0 \rangle $ denoted $|3 \rangle$. A $750$~mG
magnetic field oriented along the beam propagation direction lifts
the degeneracy of the Zeeman levels in the $F=2$ and $F=3$
manifolds.
Atoms which are not in the $m_F=0$ state are
out of Raman resonance and make no
transitions. 
The Raman beams
propagate at $43^{\circ}$ with respect to the evanescent wave
propagation vector in the $xz$-plane. It is therefore
necessary to rotate the magnetic field adiabatically while the
atoms fall to the mirror in order that they be in an
eigenstate of the
polarization (nearly circular) of the evanescent wave. After
reflection the magnetic field is turned back to coincide with the
propagation direction of the Raman beams.

To generate the Raman beams (separated by
$\omega_{23}/2\pi=3.036$~GHz), we modulate the injection current of a
free running diode laser at $1.5$~GHz and inject the $\pm 1$
sidebands into two slave laser diodes (each diode is injected by one
sideband). Before injection, the laser beam passes through a
Fabry-Perot cavity of $3$~GHz spectral range, a finesse of $130$, and
locked to the two sideband wavelengths to filter out the carrier
wavelength and the $\pm 2$ sidebands. After injection, each Raman
beam passes through an acousto-optic modulator (AOM). An arbitrary
function generator modulates these AOMs to produce Blackman pulses
~\cite{Kasevich:92a} of the desired duration.
Observation of the beat note between the two slave laser beams
indicates a relative frequency spread less than $20$~Hz (equal to the
resolution bandwidth of the spectrum analyser) sufficiently narrow
that the Raman transition width is not limited by the Raman phase
coherence.

To test our setup, we first make a velocity selection and
immediately analyse it with a second Raman pulse. At $t=0$, we
prepare the atoms in the $F=3$ state. At $t=8$~ms we apply a
Raman $\pi$ pulse with a detuning $\delta_{\rm{S}}$ to transfer
atoms to $|2 \rangle$. Then comes a pushing beam resonant with the
$5S_{1/2}F=3 \longrightarrow 5P_{3/2}F'=4$ transition which
removes all the atoms remaining in $F=3$. At $t=22$~ms we apply a
second Raman ``analysis" $\pi$ pulse with a detuning
$\delta_{\rm{A}}$ to transfer atoms back to $|3 \rangle$. The
atoms in the $F=3$ level are detected via the fluorescence induced
by a retroreflected probe laser resonant with the $5S_{1/2}F=3
\longrightarrow 5P_{3/2} F'=4$ transition and collected in a
$0.1$~sr solid angle on a photomultiplier tube. No repumper is
present in order to avoid detection of atoms in the $F=2$ level.

\begin{figure}
\begin{center} \
 \includegraphics{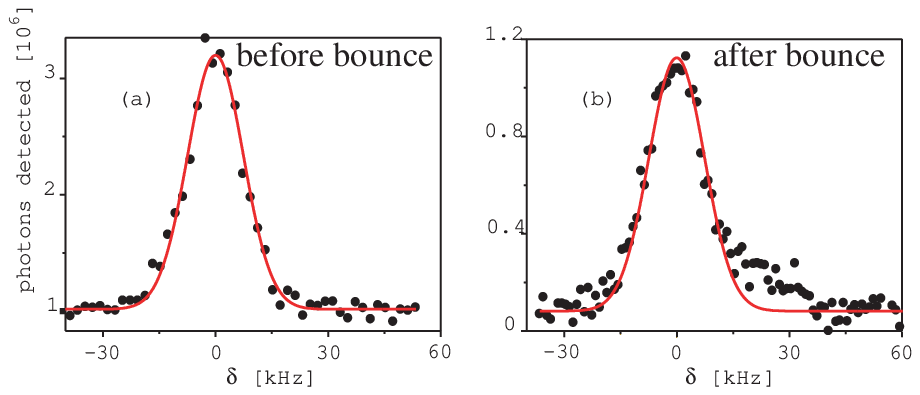}
\end{center}
\caption{Transverse atomic velocity distribution before 
(a) and after reflection (b) with $\Delta_{\mathrm{EW}}=2.4$~GHz.
We scan the detuning $\delta$ of the second Raman pulse. 
The solid line of
(a) is a Gaussian fit to the data (the half width
at $1/\sqrt{e}$ corresponds to
$0.47$~$v_{\rm{rec}}$). 
In (b), in addition to the data, 
we have plotted the same Gaussian as
in (a), normalized to the height of the data in (b)
in order to emphasize the presence of a pedestal. 
}
\label{fig.specularite}
\end{figure}

We repeat the sequence with a different value of $\delta_{\rm{A}}$ in
order to acquire the transverse velocity distribution of incident
atoms (Figure~\ref{fig.specularite}(a)). The half width at $1/
\sqrt{e}$ of the distribution is $7.3$~kHz. 
This width is consistent with what is
expected for a $150$~$\mu$s Blackman pulse. The curve demonstrates a
velocity selection width for a single pulse of $0.33$~$v_{\rm{rec}}$
(HW at $1/ \sqrt{e} $) along the propagation direction of the Raman
beams. This is about $20$ times narrower than the velocity width in
the MOT. Because the analysis sequence has the same resolution as the
selection sequence, our velocity resolution is $\sqrt{2}$ times
larger, that is $0.47$~$v_{\rm{rec}}$. This resolution is $3$ times
better than what was used in Ref.~\cite{landragin:96a}.

To observe the effect of the reflection on the transverse velocity, 
we proceed in a manner analogous
to that described above (see Fig.~\ref{fig.disporaman}).  At $t=0$ we 
prepare the atoms in
the $F=2$ level. The Raman selection pulse transfers a narrow
velocity class to $|3 \rangle$ at $t=8$~ms. The atoms then fall onto
the mirror. The frequency of the evanescent wave is tuned to the blue
of the $F=3$ resonance to the excited state of the $D_2$ line, and to
the red of the $F=2$ resonance. Atoms in $F=2$ do therefore not
reflect from the mirror. After reflection ($t=120$~ms), the
analysis pulse transfers some atoms back into $|2 \rangle$. Next the
pushing beam removes the atoms remaining in the $F=3$ level and
finally we detect the atoms transferred to $|2 \rangle$ with the
probe laser and the repumper.

\begin{figure}
\begin{center} \
\includegraphics{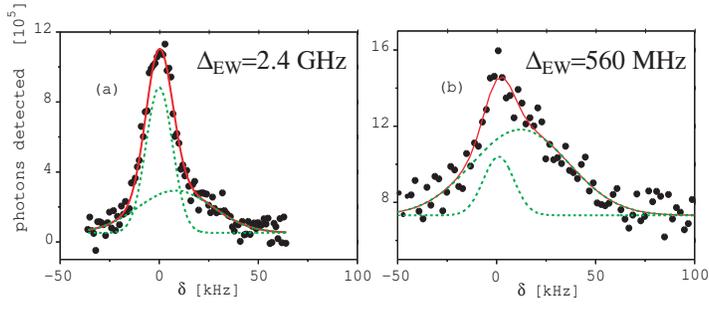}
\end{center}
\caption{Atomic velocity distribution for two different values of
$\Delta_{\mathrm{EW}}$.
The solid 
lines show a fit
using two Gaussian curves as described in the text.
Both the individual Gaussians as well as their sum are shown.
Each atom results in about 100 detected photons.
} 
\label{fig.fit}
\end{figure}

Atoms which have not been selected by the first Raman selection
pulse can contribute to a background if they happen to be pumped
into the $|3 \rangle$ state (by the evanescent wave, for example)
during their trajectory. We measure this background using the same
sequence described just above with the selection detuning
$\delta_{\rm{S}}$ tuned far from resonance. In our data
acquisition we alternate between normal and background
measurements and subtract the background on each shot. The
detuning of the analysis pulse $\delta_{\rm{A}}$ is scanned
randomly over the desired values, and we acquire and average about
$3$ measurements for each value of $\delta_{\rm{A}}$ to acquire a
spectrum such as that shown in Fig.~\ref{fig.specularite}(b). The
peak value in Fig.~\ref{fig.specularite}(b) corresponds to about
$10^4$  atoms detected per bounce. Typical values of the background 
in this 
case correspond to about $3 \times 10^3$ atoms.
Despite this subtraction, we observe a non-zero background in
Figs.~\ref{fig.specularite} and \ref{fig.fit}. 
This background appears to be due to atoms which 
reflect from the mirror but are pumped into the $F=2$ state
after reflection.

With this system, we first checked that the atoms obey 
the law of reflection, that is the reflected angle is equal to
the incident one. 
We vary the mean velocity of the initial distribution by 
choosing an appropriate $\delta_{\mathrm{S}}$,
and verify that the center of the reflected
velocity distribution varies by the same amount. 
We have noted a non-zero intercept
of this linear dependence which we attribute to
a slight tilt (about $1^{\circ}$) in the mirror
relative to the horizontal.

For a reflection to be regarded as truly specular, 
the velocity distribution must remain unchanged after reflection.
Figure~\ref{fig.specularite}(b) shows the velocity distribution
(the number of atoms detected in the $F=2$ state after the Raman
analysis pulse) after the bounce. One distinguishes a narrow peak
whose width appears identical to the initial one, and a broad
pedestal whose center is shifted by 7.9 kHz, an amount corresponding 
to a
$0.5 \hbar k_{\rm{L}}$ momentum transfer with respect to the narrow one
along the observation direction. This transfer is in the same
direction as the evanescent
wave propagation vector, and remains so when the evanescent wave 
(Ti:S) laser
direction is reversed (that is it also reverses). 

\begin{figure}
\begin{center} \
\includegraphics{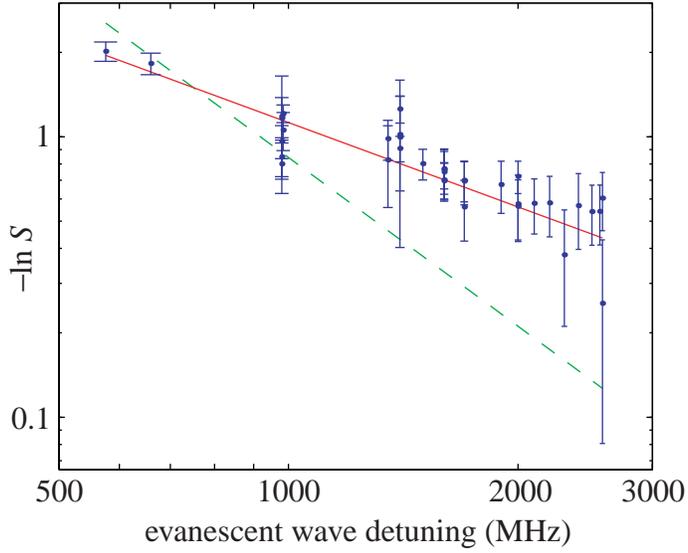}
\end{center}
\caption{Variation of $S$ the fraction of atoms in the narrow 
peak as a function of the evanescent wave detuning 
$\Delta_{\mathrm{EW}}$.
The solid line (slope $-1$) corresponds to 
$S=\exp{-\alpha/\Delta_{\mathrm{EW}}}$.
The dashed line (slope $-2$) corresponds to 
$S=\exp{-\beta/\Delta_{\mathrm{EW}}^{2}}$.
Here $\alpha$ and $\beta$ are fit parameters.
}
\label{fig.loidechelle}
\end{figure}

In an attempt to understand the origin of the pedestal, we
acquired several reflected velocity distributions under differing
conditions; two examples are shown in Fig.~\ref{fig.fit}. Each
such distribution is fitted to a sum of two Gaussians plus
a flat background. 
We first
examined the parameters of the pedestal as a function of the
evanescent wave 
detuning $\Delta_\mathrm{EW}$. 
We observed little variation of the width and the shift 
relative to the narrow peak. 
To simplify the study of the relative contribution of the
two components, 
we fixed the width of the narrow peak at the measured width of 
the resolution function.
We also fixed the width of
the pedestal and the shift at the average values of our preliminary 
fits:
the pedestal width was fixed to be that of the convolution of our 
resolution function
and a Gaussian of 18 kHz rms and 
the shift to be 7.9 kHz.
Using this analysis we can measure the fraction $S$ of atoms detected
in the narrow peak 
as a function of $\Delta_{\mathrm{EW}}$.
(See Fig.~\ref{fig.loidechelle}.)
The data are well fit by $S=\exp (-\alpha/\Delta_{\mathrm{EW}})$
with $\alpha=1.1$~GHz. 

The above detuning dependence immediately suggests spontaneous
emission within the evanescent wave which reduces the numbre of 
specluarly reflected atoms by a factor of $\exp{(-N_{\rm{SE}})}$.
A simple estimate of $N_{\rm{SE}}$, 
the average number of spontaneous emissions,
is given by $N_\mathrm{SE}=2\pi/(\lambda_{\mathrm{dB}} 
\kappa)\times\Gamma/\Delta_{\mathrm{EW}}$, 
where $\Gamma/2 \pi=5.9$~MHz is the natural linewidth of the atomic transition
~\cite{Kasevich:90a,voigt:00a}. 
A better estimate 
includes the modification of the
potential due to the van der Waals
interaction~\cite{Landragin:96b}, 
the modification of
the spontaneous emission rate close to the
surface~\cite{Courtois:96a},
and an average of these effects over the mirror surface.
We find, in our range of detunings, 
that $N_{\rm{SE}}$ still varies as $\Delta_{\mathrm{EW}}^{-1}$
to a good approximation
but with a probability about 1.5 times higher.
To calculate $S$ one must
also take into account
the fact that at large detunings the branching ratio
for falling back into the $F=3,m_{F}=0$ state, the only 
one which we detect, is 2/3. 
This factor cancels the increase in $N_{\rm{SE}}$
due to the effects of the dielectric surface.
One predicts therefore $S=\exp 
(-\alpha_{\rm{SE}}/\Delta_{\mathrm{EW}})$
with $\alpha_{\rm{SE}}=0.55$~GHz. 
There appears to be too little spontaneous emission 
(by a factor of 2)
to entirely explain our results. 

In addition, spontaneous emission in the
evanescent wave should result in 
an average momentum transfer of $\hbar k_{x}$ along $x$,
that is,
a shift of the broad pedestal
of $\hbar k_x \times \cos (43^{\circ}) = 1.1 \hbar k_{\rm{L}} $ along the
Raman beam direction instead of the $0.5 \hbar k_{\rm{L}}$ that we
observe. 
This observation confirms the above conclusion
that spontaneous emission in the evanescent wave is only partly 
responsible for our observations.
This is in contrast to the study of Ref.~\cite{voigt:00a}
which used a very small value of $\kappa$ to
get a large number of spontaneous emissions in the
evanescent wave. 

Another mechanism which causes diffuse reflection is
discussed in Ref.~\cite{Henkel:97b}.
It involves scattered Ti:S light which
interferes with the evanescent wave. The interference produces a
rough potential which diffusely scatters the
atoms.
This mechanism does {\it not} involve spontaneous 
emission. 
The scattered light could come either
from the surface roughness, inhomogeneities in the bulk of the
prism or some other object such as a prism edge. 
Using the results in
Ref.~\cite{Henkel:97b} one can show that the propagating modes of
the scattered light would cause $S$ to
vary as $\exp (-\alpha_{\rm{R}}/\Delta_{\mathrm{EW}})$.
Where $\alpha_{\rm{R}}$ depends on the amount of scattered light.
The pedestal due to this effect should exhibit no shift
relative to the specular peak.
Since spontaneous emission is responsible for about
one half the pedestal, we expect
a pedestal shifted by about one half of 
the shift due to spontaneous emission alone
in good agreement with our observations. 
Since both effects have the same $\Delta_{\rm{EW}}$ dependence
the observed shift should not depend on the detuning.

Another possible explanation for the pedestal is
spontaneous emission induced by the stray
light above the prism while the atoms fall towards the mirror. 
Indeed, in experiments in which we left
the evanescent wave laser on for $40$~ms while the atoms fell towards 
the
mirror, we observed an optical pumping of the atoms from their initial
hyperfine state (F=3) to the other hyperfine state (F=2)
(about $10\%$ of the atoms were lost in this way at 
$\Delta_{\mathrm{EW}}=940$~MHz).
This mechanism predicts that $S$ should vary as
$\exp -\beta/\Delta_{\mathrm{EW}}^2$, where $\beta$ is a constant 
which
depends on the mean light intensity experienced by the atoms.
As shown in Fig.~\ref{fig.loidechelle}
the data are not consistent with this dependence.

Thus we believe that we have identified the source of the 
diffuse reflection in our experiment as
the sum the effects of the scattering
of atoms from a potential induced by the random 
interference pattern of
the evanescent wave and stray light,
and spontaneous emission in the evanescent wave.
According to this interpretation
we have $1.1\:\rm{GHz} = \alpha_{\rm{R}}+\alpha_{\rm{SE}}$ 
and we can work out the effective mirror 
roughness associated with $\alpha_{\rm{R}}$. 
We find $\sigma\approx 0.3$~nm, a value much larger than the prism's 
measured surface roughness (0.07 nm). Since the effective mirror 
roughness due to light scattering by the prism surface is of the 
order of the surface roughness itself~\cite{Henkel:97b}, we presume 
that most of the stray light is from other sources.

We turn now to an analysis of the narrow peak. 
Since the area under the
broad peak can be reduced by increasing the
detuning $\Delta_{\mathrm{EW}}$, the essential question is ``How
faithfully is the initial velocity distribution reproduced in the
narrow peak?''.
To characterize this effect, we
compare the width of the narrow peak to that of the
resolution function (atomic experimental velocity distribution
before the bounce) for 36 runs acquired at different
values of $\Delta_{\mathrm{EW}}$.
We now fit
the experimental curves by a sum of two Gaussians with all
parameters adjustable
except for the width and center of the broad peak.
Averaging over 36 measurements, we find
$\sigma_{\rm{meas}}^2-\sigma_{\rm{res}}^2
=-(0.13v_{\rm{rec}})^2 \pm (0.08v_{\rm{rec}})^{2}$,
where $\sigma_{\rm{res}}$ and $\sigma_{\rm{meas}}$ 
are the
half widths at $1/\sqrt{e}$ of the two curves
after the bounce, and the uncertainty is
the standard deviation of the weighted mean of our 36 measurements.
A negative sign in the result is not necessarily unphysical
because it could be due for example, to
a slightly concave reflecting surface
which collimates the atoms.
We do not consider this deviation from zero
to be statistically significant, however.
We conclude that the observed reflection is
consistent with a specular reflection to within about $0.1\: 
v_{\rm{rec}}$.
Our limit is a factor of 10 better
than our previous best result 
\cite{landragin:96a}.

To compare our results with those of Refs.~\cite{saba:99a} and
~\cite{lau99a}, we calculate the 
rms angular deviation of an effective reflecting surface from 
perfectly
plane: $\sigma_{\theta}=\frac{1}{2}
\frac{v_{\rm{rms}}}{v_{\rm{in}}}$, where $v_{\rm{in}}$ is the
incident atomic velocity on the mirror and $v_{\rm{rms}}$ is the rms
transverse velocity added by the mirror.
Using the upper limit $v_{\rm{rms}} < 0.1\: v_{\rm{rec}}$,
we find that the
effective mirror surface is flat to within
uncertainty of $0.5$~mrad. 


Thus, we conclude that at sufficiently large detunings,
it is possible to produce a highly specular mirror for
atomic de Broglie waves. 
By analogy with photon
optics, the double structure we observe
suggests that we are in the regime where
the roughness of the atomic mirror is small compared to the
wavelength of the reflected matter wave. 
In that regime, the specular peak corresponds to a
``perfectly'' coherent reflection, and it should be possible to test
this property in an atom interferometer. 
Interferometric experimental studies are in progress.

\section*{Acknowledgments}
We acknowledge the assistance of C. Aussibal and thank
C. Henkel and J. Thywissen for useful discussions.
This work
was supported by the European Union under grants IST-1999-11055, and
HPRN-CT-2000-00125, and by the DGA grant 99.34.050.


\newcommand{\noopsort}[1]{} \newcommand{\printfirst}[2]{#1}
  \newcommand{\singleletter}[1]{#1} \newcommand{\switchargs}[2]{#2#1}

\newpage

\end{document}